\documentclass[prl,12pt]{revtex4}

\usepackage[english]{babel}
\usepackage[intlimits]{amsmath}
\usepackage[dvips]{graphicx}

\begin{document}

\title{Coupling of Two Motor Proteins: a New Motor Can Move Faster}
\author{Evgeny B. Stukalin, Hubert Phillips III, and Anatoly B. Kolomeisky}

\affiliation{Department of Chemistry, Rice University, Houston, TX 77005 USA}

 \begin{abstract}
We study the effect of a coupling between two motor domains  in highly-processive  motor protein complexes. A simple stochastic discrete model, in which the  two parts of the protein molecule  interact through some energy potential, is presented. The exact analytical solutions for the dynamic properties of the combined motor  species, such as the velocity and dispersion, are derived in terms of the properties of free  individual motor domains and the interaction potential. It is shown that the coupling between the motor domains can create a more efficient  motor protein  that can move faster than individual particles. The results are applied to analyze the motion of helicase RecBCD molecules.

\begin{flushright}
\text PACS numbers: 82.39-k, 82.39Fk, 87.15Aa
\end{flushright}

\end{abstract}

\maketitle

Motor proteins  are active enzyme molecules that are responsible for generation of forces and molecular transport in biological systems \cite{howard_book, bray_book}. They move along polar molecular tracks such as cytoskeletal filaments and DNA molecules, and this motion is powered  by  energy released from a hydrolysis of adenosine-triposphate  (ATP) molecules or related compounds.  However, the mechanisms of the chemical energy  transformation into the mechanical work are not fully understood \cite{howard_book}.

Crystal structures of motor proteins reveal that they can be viewed as complex systems consisting of many functional domains \cite{howard_book,singleton04}. It is assumed that this complexity of motor proteins appeared during the evolution  as a way to perform simultaneously many biological functions and to increase  efficiency. A good example is a RecBCD enzyme that belongs to a class of helicase motor proteins \cite{lohman96, tuteja04}. It processes DNA ends resulting from the double-strand breaks and other defects \cite{singleton04}. This motor protein  unwinds the DNA molecule into two separate strands, and then it digests them  by moving at the same time  along the DNA  until reaching a special signal sequence of bases, called the recombinational hot-spot (Chi site) \cite{bianco01,dohoney01,taylor03, dillingham03,spies03}. RecBCD is a hetero-trimer made of three proteins: RecB (134 kDa), RecC (129 kDa), and RecD (67 kDa)  \cite{finch1, finch2, finch3}. Two subunits,  the individual RecB and RecD domains, have helicase activities, consume ATP and act on 3'- or 5'-ended DNA strands, respectively  \cite{taylor03, dillingham03, ganesan92}. Meanwhile, the third subunit (the domain RecC)  has no ATPase activity, and  it functions as a clamp preventing the dissociation of the enzyme complex  from the track \cite{julin98}, and it also recognizes the recombinational hot-spots  \cite{singleton04}. Recent experiments provided data on DNA unwinding rates by RecBCD enzyme as well as by its active subunits  RecB  and RecD  at the single-molecule level \cite{bianco01,taylor03, bianco00,perkins04}.  Surprisingly, the speed of the protein complex is significantly faster than the unwinding rates of individual motor proteins \cite{taylor03}. In addition, RecBCD has a very high processivity, it can move more than 20000 bases without detachment, while the individual RecB and RecD subunits cannot travel more than 100 bases \cite{bianco01,taylor03,dillingham03}. These observations raise the general fundamental question important for all biological systems:  how the interaction between the different subunits is optimized to produce  highly-efficient and multi-functional molecular biological nanomachines? The work presented here aims to address this specific issue by developing a simple stochastic model of the motion of motor proteins consisting of two interacting subunits.  

We assume that the motor protein complex consists of two  particles, as shown in Fig. 1. Each subunit can move only along its own one-dimensional track that corresponds to the motion of RecB and RecD domains on the separate DNA strands. The positions of the particle $A$ on the upper lattice is given by the integer $l$, while $m$ specifies the position of the lower domain $B$. Due to the link between the motor subunits only the limited number of the motor complex configurations has to be considered. In the simplest description, we assume that 3 configurations are possible, i.e., $0 \le | l-m | \le 1$ (see Fig. 1). More configurations and different interactions can be easily accounted in the model.  Our approach is  related to the theoretical model of a helicase motion proposed by Betterton and J\"{u}licher \cite{julicher03}. They view  the DNA unwinding as a result of interaction between the whole helicase and DNA fork (a junction between double-stranded and single-stranded segments), while we discuss the effect of the molecular structure and the internal coupling of the subunits on the motion of motor proteins.

The dynamics of the system can be described by a set of transition rates for domains $A$ and $B$ for discrete steps forward and backward along the corresponding  tracks, as illustrated in Fig. 1. These rates depend not only on the type of the subunit, but also on the specific configuration of the cluster. However, the dependence of rates on DNA sequence is neglected in this approximation.  For  the configurations  $(l-1,l)$  the upper subunit $A$ can only move forward with the rate $u_{a1}$, while the lower particle $B$ can only hop backward with the rate $w_{b2}$: see Fig. 1a. Similarly, in the configurations $(l+1,l)$ [Fig. 1c] the domain A can only go back with the rate $w_{a2}$, while the domain $B$ can advance with the rate $u_{b1}$. In the configurations   $(l,l)$ [Fig. 1b] both particles can move forward and backward with the rates $u_{a2}$ ($u_{b2}$) and $w_{a1}$ ($w_{b1}$) for the upper (lower) subunit.

The interaction between the protein domains is specified by the parameter  $ \varepsilon \ge 0$  defined as energy difference between the states $(l \pm 1,l)$ and $(l,l)$, respectively (Fig. 1). We assume that the configuration $(l,l)$ is energetically most favored, while the configurations $(l \pm 1,l)$ have a higher energy. The possible reason for this might be the internal stress and/or the work needed to break the bond between the bases in DNA. It allows us to obtain the detailed balance thermodynamic relations for the transition rates:
\begin{equation}\label{det.bal1}
 \frac{u_{j1}}{w_{j1}} = \frac {u_{j}}{w_{j}}exp(+ \varepsilon /k_{B}T),
\end{equation}
\begin{equation}\label{det.bal2}
\frac{u_{j2}}{w_{j2}} = \frac {u_{j}}{w_{j}}exp(- \varepsilon /k_{B}T),
\end{equation}
with $j=a$ or $b$. The rates $u_{j}$ and $w_{j}$ are the forward and backward transition rates for the subunit $A$ ($j=a$) and $B$ ($j=b$) in the case of no inter-domain interaction, i.e., $\varepsilon = 0$.

We introduce $P(l,m;t)$ as the probability to find the system in the configuration $(l,m)$ at time $t$. It can be determined by  solving of a set of independent Master equations,
\begin{eqnarray}
& & \frac{d P(l-1,l;t)}{{dt}} = u_{b2} P(l-1,l-1;t) + w_{a1} P(l,l;t) - (u_{a1} + w_{b2}) P(l-1,l;t), \\
& & \frac{d P(l+1,l;t)}{{dt}} = u_{a2} P(l,l;t) + w_{b1} P(l+1,l+1;t) - (u_{b1} + w_{a2}) P(l+1,l;t). 
\end{eqnarray}
The corresponding  equation for $P(l,l;t)$ is just a linear combination of two equations presented above, and therefore it is not considered. In addition, the probabilities satisfy the normalization condition,
\begin{equation}
\sum\limits_{l = - \infty }^{ + \infty } \left ( {P(l,l;t) + P(l-1,l;t) + P(l+1,l;t)} \right ) = 1, \quad {\text{ (all }}t{\text{).}}
\end{equation}
The solutions of Master equations can be found be summing over all integers $ -\infty < l < +\infty $. Define new functions
\begin{equation}
 P_{0}(t) = \sum\limits_{l = - \infty}^{ + \infty } {P(l,l;t)}, \quad P_{1}(t) = \sum\limits_{l = - \infty}^{ + \infty } {P(l-1,l;t)}, \quad P'_{1}(t) = \sum\limits_{l = - \infty}^{ + \infty } {P(l+1,l;t)}.
\end{equation} 
Then, using the conservation of probability,  the steady-state distribution can be easily derived,
\begin{equation}
P_{0} = 1/ \Omega, \quad P_{1} = \beta/ \Omega, \quad P'_{1} = \alpha/ \Omega,
\end{equation}
where $\Omega = 1 + \alpha + \beta$, and $\alpha$ and $\beta$ are given by
\begin{equation}
\alpha = \frac{u_{a2} + w_{b1}}{u_{b1} + w_{a2}}, \quad  \beta = \frac{u_{b2} + w_{a1}}{u_{a1} + w_{b2}}.
\end{equation}

From the knowledge of the  probability densities the dynamic properties of the motor complex, such as the mean velocity $V$ and the dispersion $D$, assuming the size of the step is one, can be  calculated \cite{FK99,kolomeisky01}. The velocity is given by
\begin{equation}\label{eq.formal_vel}
V=\frac{1}{2} \left \{ (u_{a1}-w_{b2}) P_{1} + (u_{a2}+u_{b2}-w_{a1}-w_{b1}) P_{0}+(u_{b1}-w_{a2}) P'_{1} \right \},
\end{equation}
that yields the following result,
\begin{equation}\label{eq.velocity}
V  = \frac{1}{\Omega}(u_{a2} + u_{b2} - \alpha w_{a2} - \beta w_{b2}).
\end{equation}
Similarly for the dispersion we obtain
\begin{equation}
D = \frac{1}{\Omega}\label{eq.dispersion}
\left\{ {\frac{1}{2}[u_{a2} + u_{b2} + \alpha w_{a2}  + \beta w_{b2}] - \frac{{(V + w_{a2} )(u_{a2} - \alpha V)}}{{u_{b1} + w_{a2}}} 
- \frac{{(V + w_{b2})(u_{b2}  - \beta V)}}{{u_{a1}  + w_{b2} }}} \right\}.
\end{equation}
If the motor domains are identical ($A=B$), then the transition rates are related as
\begin{equation}
u_{a1}=u_{b1}=u_{1}, \quad u_{a2}=u_{b2}=u_{2}, \quad w_{a1}=w_{b1}=w_{1}, \quad w_{a2}=w_{b2}=u_{1}.
\end{equation} 
Then the expressions for the mean velocity and dispersion can be written in a much simpler form,
\begin{equation}\label{eq.vel1}
V  = \frac{2(u_{1}u_{2} - w_{1}w_{2})}{u_{1} + w_{2} + 2(u_{2} + w_{1})};
\end{equation}
\begin{equation}\label{eq.disp1}
D  = \frac{u_{1}u_{2} + w_{1}w_{2} - V^{2}}{u_{1} +w_{2} + 2(u_{2} + w_{1})}.
\end{equation} 

Now consider the case of symmetric motor domains without interaction ($ \varepsilon = 0 $). If the particles  are  not  connected in the cluster and  allowed to move freely along its tracks, then their dynamic properties are
\begin{equation}
V_{0}=u-w, \quad D_{0}=(u+w)/2,
\end{equation}
with $u_{1}=u_{2}=u$ and   $w_{1}=w_{2}=u$. However, the average velocity and dispersion of the motor protein cluster (without interaction) are different. From Eqs. (\ref{eq.vel1}) and (\ref{eq.disp1}) we obtain 
\begin{equation}
\frac{V(\varepsilon = 0)}{V_{0}} = \frac{2}{3}, \quad \frac{1}{3} \le \frac {D(\varepsilon = 0)}{D_{0}} \le \frac{10}{27}.
\end{equation}
These results show that in the case of no inter-domain coupling the speed of the combined cluster is smaller than the rates of the free particles, as expected,  although the  fluctuations are also decrease. 

Much more interesting is the situation when the two motor domains interact with each other ($\epsilon > 0$). Using the detailed balance conditions (\ref{det.bal1}) and (\ref{det.bal2}) the transition rates can be written in the following form
\begin{equation}\label{eq.rates1}
  u_{j1} = u_{j} \gamma^{1 - \theta_{j1}}, \quad   w_{j1} = w_{j} \gamma^{- \theta_{j1}}
\end{equation}
\begin{equation}\label{eq.rates2}
 u_{j2} = u_{j} \gamma^{- \theta_{j2}}, \quad  w_{j2} = w_{j} \gamma^{1 - \theta_{j2}} 
\end{equation}
where $\gamma = exp(\varepsilon/k_{B}T)$, and $j=a$ or $b$. The coefficients $\theta_{ji}$ determine how the interaction energy is distributed between the forward and backward transitions \cite{howard_book,julicher03,FK99}. They are closely related to the load-distribution factors that have been used successfully in the stochastic single-particle models of motor proteins \cite{FK99}. It is reasonable to approximate the distribution factors $0 \le \theta_{ji} \le 1$ as equal to each other since they describe the energy of the similar processes in the motion of individual motor domains. However, the situation of the state-dependent  energy-distribution factors can also be considered.  

For the motor protein complex  with the symmetric domains ($A=B$) the effect of interactions can be analyzed by looking at the ratio of the cluster velocity to the velocity of the free non-interacting particles, 
\begin{equation}
r_{V}=\frac{V}{V_{0}} = \frac{2 \gamma^{1 - \theta}}{2+ \gamma}.
\end{equation}
The dependence of the relative velocity on the interaction energy for different  $\theta$ is shown in Fig. 2. The most interesting observation is  that for small values of the energy distribution factors ($\theta \le 0.23$) there is a range of interaction energies when the velocity of the motor protein's complex is {\it faster} than the velocities of the free particles. It contradicts the naive intuitive expectations, but it can be understood by considering again Eqs. (\ref{eq.formal_vel}), (\ref{eq.rates1}) and (\ref{eq.rates2}). Large interaction energies increase the transitions rates $u_{j1}$ and  $w_{j2}$, and lower the rates  $u_{j2}$ and  $w_{j1}$  for $j=a$ or $b$ as indicated in Eqs. (\ref{eq.rates1}) and (\ref{eq.rates2}).  At the same time the probabilities of non-vertical configurations $P_{1}$ and $P_{1}'$ are exponentially decreasing functions of the interaction energy. Then each term in Eq. (\ref{eq.formal_vel}) has a maximum at some specific value of the interaction energy. Thus the dependence of the relative velocity on the interaction energy is a result of two opposing factors: the acceleration of  some forward  transition rates is balanced by the decrease of the probabilities of the configurations from which the motion is possible.  In the limit of very large interactions the cluster will not move since it could only be found in the vertical configurations.

The expression for the relative  dispersion of the motor protein cluster with symmetric domains ($A=B$) is given by
\begin{equation}
r_{D}=\frac{D}{D_{0}} = \frac{2 \gamma^{1 - \theta}}{2+ \gamma} \ g(u,w;\gamma)=\frac{2 \gamma^{1 - \theta}}{2+ \gamma}  \ \left[ 1- \frac{2u w}{(u+w)^{2}} -\frac{4 \gamma}{(2+\gamma)^{2}} \left(\frac{u-w}{u+w} \right)^{2} \right ],
\end{equation}
where it can be shown that $0.5 \le g(u,w;\gamma) < 1$.  The interaction energy also changes the dispersion in the similar way as the velocity, however, the effect is smaller. It can be seen from Fig. 3 where the ratio of the relative dispersion to the relative velocity is plotted for different transition rates. For all values of the parameters the relative dispersion is always smaller than the relative velocity. It is interesting to note that there are situations when $r_{V} >1$ and  $r_{D} <1$, i.e., the motor complex  moves faster but fluctuates less than the free subunits, making it an extremely efficient motor protein. 

We can now apply our model for the analysis of the motion of RecBCD helicases. It was shown  experimentally that at 37 ${}^o$C and 5 mM ATP  the mutant RecBCD* (the domain RecD is nonfunctional) and the mutant RecB*CD (the domain RecB is nonfunctional) unwind the DNA with rates 73 and 300 nucleotides/s, correspondingly \cite{taylor03}. Electron microscopy and biochemical assay data indicate that the free helicases RecB and RecD move with the same speeds as the corresponding mutant RecBCD* and RecB*CD enzymes \cite{korangy94}. It can be  argued that  the backward rates $w_{ji}$ are small  \cite{perkins04}, since the backward steps are rarely seen in the experiments on helicases. Then the average speed of DNA unwinding by the RecBCD complex can be approximated as
\begin{equation}
V \approx \frac{(u_{a}+u_{b}) \gamma^{1-\theta}}{\gamma + (u_{a}/u_{b}+u_{b}/u_{a})},
\end{equation} 
where $u_{a}=73$ and $u_{b}=300$ nucleotides/s are the velocities of the free RecB and RecD proteins. Assuming $\theta \approx 0$, we obtain that $V=370$ nucleotides/s for the  interaction energy  $\varepsilon \simeq 6 \ k_{B}T$, in agreement with experimentally observed values of the RecBCD velocity \cite{taylor03}. The predicted energy of inter-domain interaction is very reasonable since it is larger than  2 $k_{B}T$ needed to break the bond between the base pair in DNA \cite{lohman96,julicher03}, but it is also smaller than a  strong covalent chemical bond energy. Note also, that the maximal possible speed of the motor protein complex cannot be larger than the sum of the velocities of the individual domains. Thus RecBCD is working with almost  maximal possible efficiency.

Our theoretical model has been greatly stimulated by the experimental observations on helicases. However, we believe that the effect of the interaction between the domains is very common for many biological systems,  especially for motor proteins. Recent experiments on KIF3A/B kinesins, that are heterodimeric processive proteins involved in intraflagellar transport and Golgi trafficking, suggest that the interaction between the motor domains is important for maximizing the performance of these enzymes \cite{zhang04}. By comparing  the properties of mutant homodimeric KIF3A/A and KIF3B/B proteins with the wild-type molecules, it was shown that the independent sequential hand-over-hand model cannot explain the heterodimer velocity data. The theoretical model similar to the one discussed above can be developed to describe the coordination between motor heads by assuming the interaction between them. We believe that taking into account the interaction between different domains is critical in construction of more realistic models of protein dynamics.  

The presented  theoretical approach neglects several features that might be important for understanding the mechanisms of biological transport phenomena. Our description of the protein dynamics and the biochemical transitions is rather oversimplified. For example, the existence of intermediate conformations and states is not taken into account \cite{FK99}. We also assumed that the transition rates depend only on the configuration of the cluster and independent of the specific nature of DNA bases. At the same time the sequence dependence is rather weak for most helicases \cite{lohman96}. In addition, the effects of DNA elasticity and inter-domain flexibility have not been considered. 

In summary,  the effect the  inter-domain coupling in  the motor protein complexes has been discussed by developing a  simple stochastic discrete model. It was shown, using the explicit analytic formulas  for the velocity and dispersion, that the interaction between different subunits in the enzyme complex might accelerate the speed of the cluster, as compared with the velocities of the free moving domains, without the significant increase in the fluctuations. This effect is due to the fact that the energy of interaction favors the compact vertical configurations of the cluster, and it  influences the forward and backward transitions differently.The asymmetry in energy-distribution factors results in a more efficient  dynamics of the motor protein complex. Our theoretical method is used successfully to describe the dynamic properties of RecBCD helicases, and it is also  relevant for other motor protein systems. Thus we present a possible explicit mechanism of the interaction between different domains in complex biological systems. It is interesting to note that the expressions for $V$ and $D$ [see Eqs. (\ref{eq.velocity}) and (\ref{eq.dispersion})] are identical to those obtained for a stochastic periodic model in which a single particle can move along two parallel tracks \cite{kolomeisky01}. It means that the motion of two coupled particles  can be mapped into the dynamics of the single particle on parallel-chain lattices. It indicates also that this approach can be generalized to include more motor protein's configurations and domains. Finally, the model can be also  extended to describe the effect of external forces on protein dynamics \cite{FK99,kolomeisky01}.

{\it Acknowledgments}. The authors would like to acknowledge the support from the Welch Foundation (grant C-1559), the Alfred P. Sloan Foundation (grant BR-4418) and the U.S. National Science Foundation (grant  CHE-0237105).

\newpage

\noindent {\bf Figure Captions:} \\\\

\noindent Fig. 1. Schematic view of a motor protein that  consists of two domains. Transition rates $u_{ai}$ and $ w_{ai}$ with $i=1$ or 2  describe the motion of the domain  $A$ (small circles), while for the second  particle  $B$ (large circles) the transitions rates are $u_{bi}$ and $ w_{bi}$. Only three  configurations are allowed: (a) $(l-1,l)$ with the energy of interaction $\varepsilon >0$; (b) $(l,l)$ with  $\varepsilon =0$; and (c) $(l+1,l)$ with  $\varepsilon >0$.

\vspace{5mm}

\noindent Fig. 2. The relative velocity for the complex motor protein with symmetric domains as a function the interaction energy for different energy-distribution factors $\theta$.

\vspace{5mm}

\noindent Fig. 3. The ratio of relative dispersion and relative velocity for the motor protein complex with symmetric domains as a function of the interaction energy for different forward and backward rates.

\newpage

\noindent \\\\\\

\begin{figure}[ht]
\unitlength 1in
\resizebox{3.375in}{4.27in}{\includegraphics{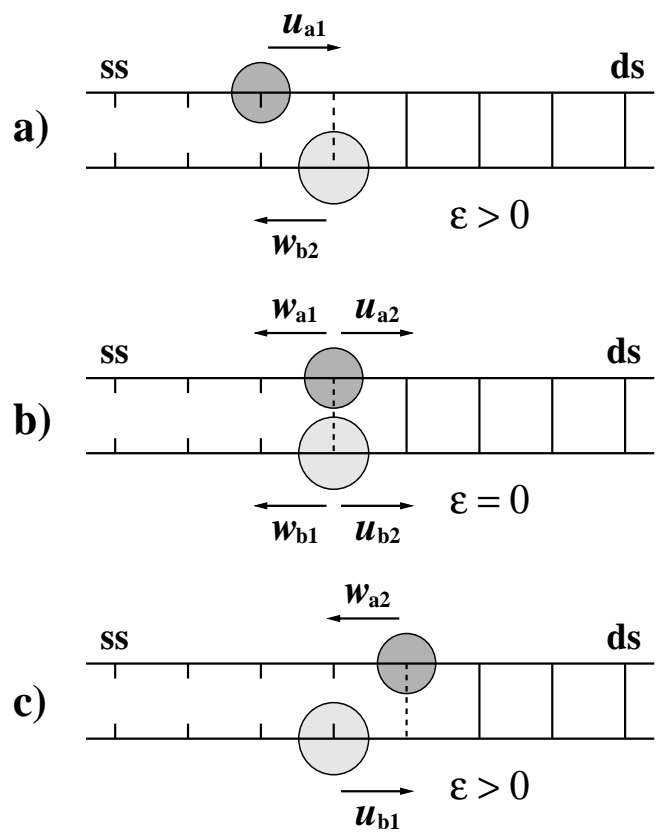}}
\vskip 0.3in
\caption{{\bf E. Stukalin, H. Phillips, A. Kolomeisky, Physical Review Letters.}}
\end{figure}

\newpage

\noindent \\\\\\

\begin{figure}[ht]
\unitlength 1in
\resizebox{3.375in}{2.50in}{\includegraphics{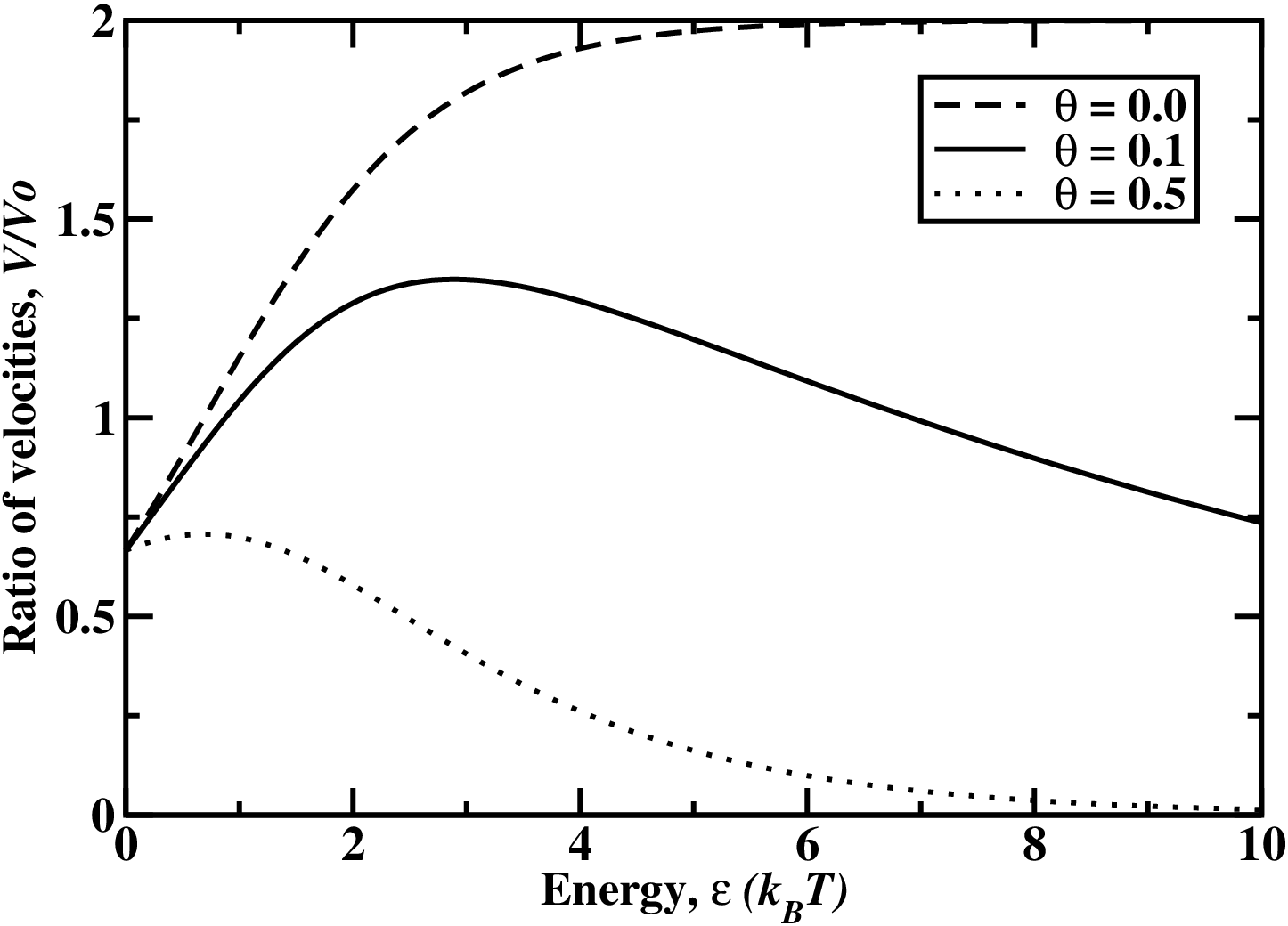}}
\vskip 0.3in
\caption{{\bf E.B. Stukalin, H. Phillips, A.B. Kolomeisky, Physical Review Letters.}}
\end{figure}

\newpage

\noindent \\\\\\

\begin{figure}[ht]
\unitlength 1in
\resizebox{3.375in}{2.50in}{\includegraphics{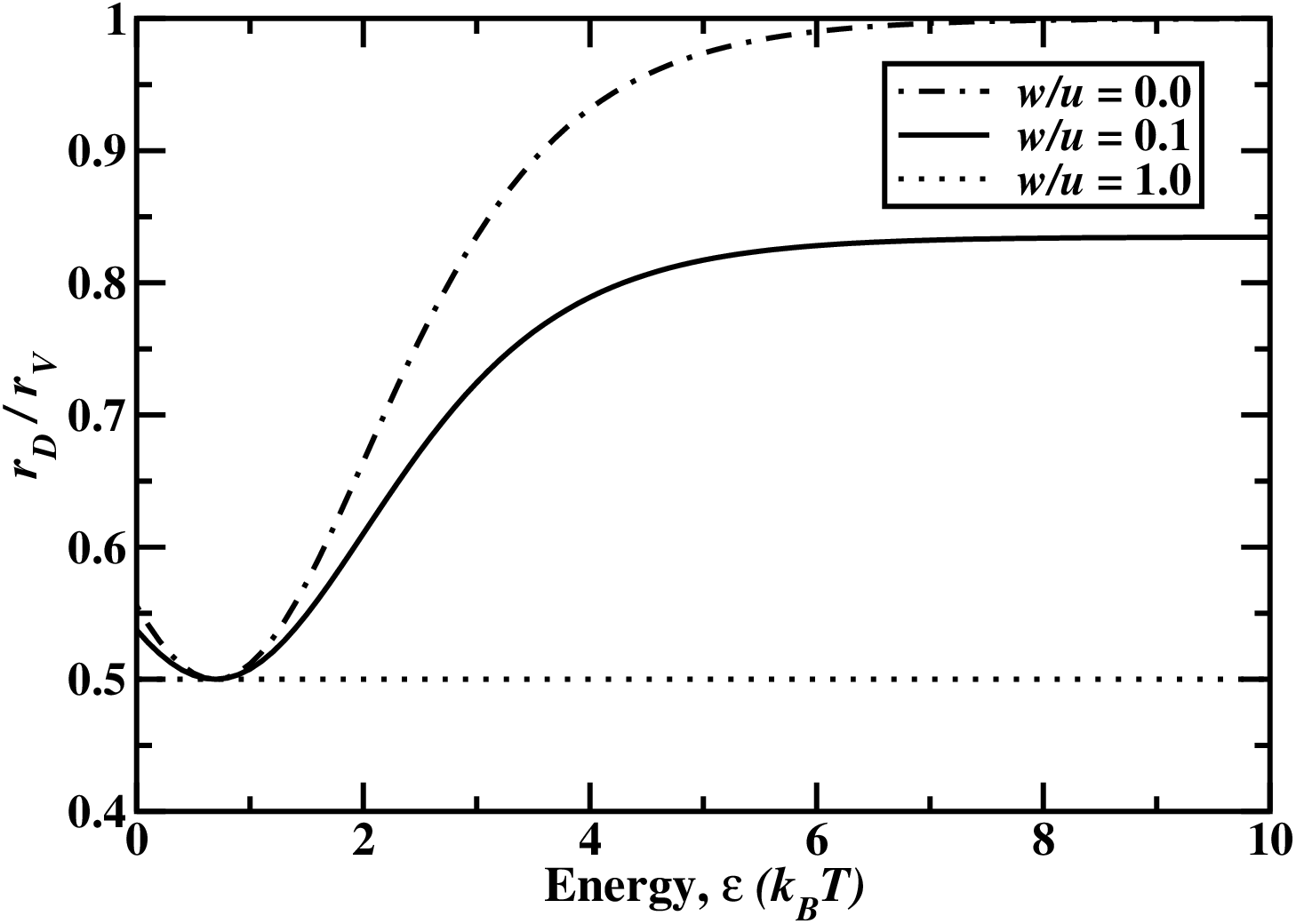}}
\vskip 0.3in
\caption{{\bf E.B. Stukalin, H. Phillips, A.B. Kolomeisky, Physical Review Letters.}}
\end{figure}

\end{document}